\documentclass[12pt]{article}
\usepackage{makeidx}
\usepackage{multirow}
\usepackage{multicol}
\usepackage[dvipsnames,svgnames,table]{xcolor}
\usepackage{graphicx}
\usepackage{epstopdf}
\usepackage{ulem}
\usepackage{hyperref}
\usepackage{amsmath}
\usepackage{amssymb}

\title{The emergence of the two cell fates and their associated switching for a
negative auto-regulating gene}

\author{Zhenlong Jiang$^{ac}$ (equal contribution), Li Tian$^{a}$ (equal contribution),\\
Xiaona Fang$^{a}$(equal contribution), Kun Zhang$^{a}$(equal contribution),\\
Qiong Liu$^{a}$(equal contribution), Qingzhe Dong$^{a}$, Erkang Wang$^{1}$, Jin Wang$^{abc*}$\\
\small{$^{a}$State Key Laboratory of Electroanalytical Chemistry, Changchun Institute of Applied}\\
\small{Chemistry, Chinese Academy of Sciences, Changchun, Jilin, 130022, China}\\
\small{$^{b}$Department of Chemistry, Physics and Applied Mathematics, State University of New}\\
\small{York at Stony Brook, Stony Brook, New York, 11794-3400, USA}\\
\small{$^{c}$College of Physics, Jilin University, Changchun, Jilin, 130012, China}\\
\small{$^*$ Corresponding author. Tel: +1-631-816-5920, Fax: +1-631-632-7960.}\\
\small{E-mail address:jin.wang.1@stonybrook.edu}}

\begin{document}
\maketitle

\section*{Abstract}

Decisions in the cell that lead to its ultimate fate are important for cellular
functions such as proliferation, growth, differentiation, development and death.
Understanding this decision process is imperative for advancements in the
treatment of diseases such as cancer. It is clear that underlying gene regulatory
networks and surrounding environments of the cells are crucial for function. The
self-repressor is a very abundant gene regulatory motif, and is often believed to
have only one cell fate. In this study, we elucidate the effects of
microenvironments mimicking the epigenetic effects on cell fates through the
introduction of inducers capable of binding to a self-repressing gene product
(protein), thus regulating the associated gene. This alters the effective
regulatory binding speed of the self-repressor regulatory protein to its
destination DNA without changing the gene itself. The steady state observations
and real time monitoring of the self-repressor expression dynamics reveal the
emergence of the two cell fates, The simulations are consistent with the
experimental findings. We provide physical and quantitative explanations for the
origin of the two phenotypic cell fates. We find that two cell fates, rather than
a single fate, and their associated switching dynamics emerge from a change in
effective gene regulation strengths. The switching time scale is quantified. Our
results reveal a new mechanism for the emergence of multiple cell fates. This
provides an origin for the heterogeneity often observed among cell states, while
illustrating the influence of microenvironments on cell fates and their
decision-making processes without genetic changes.

Keywords: gene expression $\vert{}$ self-repressor $\vert{}$ biomodality
$\vert{}$ cell fate decision making

\section*{Significance}

It is often believed that genotypes determine phenotypes. Many studies have
focused on genetic mutations rather than environmental changes or epigenetics.
Here, we design a simple self-repressing gene circuit in \textit{Escherichia
coli}. We elucidate the effects of microenvironments or epigenetics on gene
expressions through the introduction of inducers capable of binding to the
self-repressor regulatory protein. This slows down the effective binding
(regulation strength) of the regulatory protein to DNA. Despite the long-held
belief that only one cell fate is present for the self-repressor, we observe that
at some induction conditions, cells show two expression states, indicating two
cell fates. Real-time monitoring of self-repressor expression during cell growth
reveals the switching dynamics for cell fate decision-making between these two
populations.

\section*{Introduction}

Uncovering the origin of the phenotypes or fates of the cell and their
associated switching is important for the full understanding of cell functions
such as proliferation, growth, differentiation, development, and death. This
remains a challenging issue in biology. It is clear that the underlying gene
regulatory networks are crucial in determining the function of the cell (1-6),
and it is often believed that the genotype determines the phenotype (7-11).
Recently, some studies have indicated that microenvironments or epigenetics can
also alter the fates of the cell or its phenotypes even with the same genotypes
(12-18). In other words, there is a possibility that apart from mutating the
genes or the nodes themselves in the gene circuit, changing the underlying gene
regulatory wirings among the genes or nodes in the regulatory network can alter
the cell phenotypes or fates. In this study, we aim to study how altering gene
regulation determines cell fates.

Negative auto-regulation is abundant: it is found in nearly 50\% of
the feedback loops in gene regulatory networks. It is widely believed that
negative auto-regulation leads to a reduction of the gene expression noise, an
increase of gene response times, an induction of possible oscillatory gene
expression, and an improvement of the stability~of proteins produced by the
underlying gene networks (19-26). Despite these novel findings, most experimental
studies have been focused on the influences of the genetic structures themselves,
rather than the environmental or the epigenetic effects on the self-repressor.

For a self-repressing system, the expression distribution is commonly more
concentrated and well-distributed (27). Many previous investigations have reached
similar conclusions, observing only one cell fate (25, 28-30). However, these
experiments were performed mostly in simple organisms such as bacteria, for which
it is often assumed that the speed of regulatory protein binding to the
corresponding DNA for switching is significantly faster than the synthesis and
degradation of the corresponding regulatory proteins. In fact, in most organisms,
cell complexes such as the nuclei inside mammalian cells may give rise to
effectively slower processes of the underlying gene regulatory binding, due to
environmental complexities such as epigenetic effects through histone
modification or DNA methylation. That is, the effective rates of
binding/unbinding of the regulatory proteins to the DNA can be comparable to, or
even slower than, the production and degradation rate of the regulatory proteins
(31){\footnotesize}. Modeling studies (32, 33) indicate that, in this case, the
protein expressions of a negative feedback loop may not always show a simple
single steady state, but instead can show two steady states, resulting in two
different cell fates. Since the auto-regulation circuit involves only a single
gene, it is the simplest gene regulation in \textit{vivo}. We will show
experimentally that this simple gene auto-regulation circuit can lead to
different cell fates or phenotypes under specific conditions, rather than that of
only one cell fate as is commonly expected.

\section*{ Results}

\subsection*{Self-Repressing Gene Circuit and Non-Regulatory Gene Circuit}

In this study, we have designed and constructed a purely negative
auto-regulation feedback loop circuit (self-repressing gene circuit) in
\textit{Escherichia coli }(\textit{E}. \textit{coli}). The Ptet promoter
including two\textit{ tetO} operons controls the production of its repressor,
\textit{TetR}. Meanwhile, the \textit{TetR} was fused with a fluorescence protein
(Venus) for experimental measurements of the \textit{TetR} expressions. The
inducer, aTc (anhydrotetracycline), was introduced to mimic environmental
influences on expressions of the self-repression system. In the presence of an
inducer, the repressor \textit{TetR} can change its conformation and dissociate
from specific binding sequences of the DNA (T\textit{etO}). This allows for the
transcription of \textit{TetR}-Venus (Figure 1A). In order to avoid fluctuations
in copy numbers of the plasmids, the constructed circuit in the plasmid was
integrated into the chromosome of E. coli. We also constructed a series of
self-repressing circuits with different affinities to the \textit{TetR} protein
(MG::PR-WT, MG::PR-1G) (Figure S2). We chose MG::PR-8T as the main circuit of
this study for its stability and bimodal behavior.  To compare this with our
self-repressing circuit construction MG::PR-8T, we designed a non-regulatory
circuit as a control group: the MG::PR-8T-P39K circuit (Fig. 1B).

\subsection*{The Expression Distributions of the Self-Repressor Gene Circuit under
Microscopy}

To obtain the expressions of \textit{TetR} under different induction conditions,
we measured the average fluorescence signals of the reporter protein Venus for
the strain of MG::PR-8T at different inducer concentrations (300 ng/mL-1500
ng/mL) across cell populations using a wide-field fluorescence microscope. Cells
were collected and measured after being cultured in M9 medium and induced by aTc
for 4\textasciitilde{}6 hours to a logarithmic phase. To ensure accuracy of the
expression distribution, we collected no less than 10$^{3}$ cells to measure for
each sample. All expression distributions under different induction
concentrations are shown in Figure 2. The results indicate that \textit{TetR}
expression distributions vary with inducer (aTc) concentrations. Under low
inducer concentrations, the expression levels of the negative regulated gene
circuit were quite low, and this gene can be considered to be in the ``off''
state for a long time. With increased inducer concentrations, the expression
levels were significantly enhanced (Fig. 2A). From the results shown in the
microscope, we can clearly see that when inducers are added to the system, the
repressor\textit{ TetR} can no longer prevent the transcription of \textit{TetR}.
When the inducer concentrations are high enough (such as 1400ng/mL and
1500ng/mL), the steady state expression distribution can become bimodal, with two
states of low and high expression levels. Meanwhile, the percentage of the cells
in the low expression state gradually increases with the increase of the inducer
concentrations (Fig. 2A). Under high inducer concentrations, the coexistence of
both phenotypes characterized by the bimodal steady state distributions of the
fluorescence intensities can be clearly seen (Fig. 2E). When we further compare
the images in Fig. 2D and 2E, it is clear to see that one section of the cells in
Fig. 2E is brighter, while other sections were dimmer, compared to most of the
cells in Fig. 2D. As can be seen from the microscopy images, the morphologies of
the bacteria cells are not influenced by the aTc inducers at a concentration
level of 1500 ng/mL (Fig. 2E). The corresponding distributions of those images
are given in Fig.2A. In our control experiments, similar behaviors are not found
in the MG::PR-8T-P39K non-self-repressing gene circuit under the same conditions
(Figure S7). This indicates that the two expression states of \textit{TetR} were
due to the self-repressing circuit, rather than other factors such as the
influences of the inducers on the cells.

\subsection*{Fano Factor and Inhibition Curve }

To further understand our experimental observations, we need to quantify the
degrees of fluctuations. This can be measured by the Fano factor quantified as
the variance of the observable divided by the mean value (34). The Fano factor is
equal to one (\textit{F} = 1) if the distribution of the observable is exactly
Poisson. A large Fano factor implies significant statistical fluctuations
deviating from Poisson (Figure 3A). Qualitatively, the Poisson distribution
should be a good approximation for the individual ``on'' and ``off'' states when
the observed distribution of fluorescence intensity is bimodal, because each gene
state can produce proteins almost independently of gene switching. However, the
overall Fano factor for the combined probability distribution of ``on'' and
``off'' states is much larger than 1. This is because the system is close to a
two peak (Non-Poisson) distribution with different means summed together,
producing large statistical fluctuations deviating from the single Poisson
distribution. This indicates that two Poisson processes added together will not
lead to a Poisson distribution. The analysis of the coefficient of variation (CV)
in Figure S6 also illustrates this same conclusion.

Furthermore, we investigated the inhibition curve, which describes the
proportion of the bacteria with a fluorescence intensity lower than a certain
value (Figure 3B). We can see that the proportion of the gene in its inhibited
state first decreases at low concentrations of inducer (up to aTc concentration
at 1200 ng/mL) and then increases as the inducer concentration becomes higher.
More inducers introduce more interactions with the \textit{TetR} molecules. This
slows down the effective binding of the \textit{TetR} to the DNA. Therefore, the
gene has more times to be in its ``on'' state and less of a chance of being at
the inhibition state (less inhibition capability from aTc 300 ng/mL to 1200
ng/mL). More \textit{TetR} proteins will be synthesized as a result. At certain
concentrations of inducer aTc (1200 ng/mL), the number of free \textit{TetR}
molecules synthesized from the gene's ``on'' state increases, resulting in a
comparable number of \textit{TetR} molecules to aTc molecules. This will lead to
more effective regulatory binding of \textit{TetR} to DNA. Finally, there are
more chances of the gene being in its ``off'' state. We suggest that the
increasing number of the proteins produced as a result of the presence of more
inducers at this concentration range of aTc (1200-1500 ng/mL) will eventually
promote the probability of inhibition for gene switches, since more regulatory
proteins are synthesized and available for inhibition when the inducer
concentration becomes higher. At this condition, although the total protein
expression is higher with the increase of inducer concentration, it is not high
enough that the proportion of the bacteria in the inhibited state increases due
to self-repressing regulation, leading to stronger effective inhibition. At
extremely high aTc concentrations (beyond 1900 ng/mL), one expects that the
number of the available regulatory molecules becomes far beyond the one needed
for inhibition and high expression peak should dominate. However, the aTc the
toxicity from aTc as antibacterial agent to the cells becomes effective. It is
therefore not feasible to observe the healthy cell expression distribution at
this extremely high concentration of aTc.

\subsection*{The Dynamics of \textit{TetR} Expression in Real Time}

We have seen that the self-repressing circuit can give a bimodal distribution.
In order to further explore the underlying mechanism of this behavior, we
monitored the dynamics of \textit{TetR} expression in real time. We tracked cells
during their growth and division on a microscope with a FCS2 (Focht Chamber
System 2, Bioptechs) system which provides aTc continuously to guarantee the
cells growing in the right environments (continuous flow of adequate nutrients
from fresh medium (M9) through the cells on agarose pad) and avoids potential
issue of heterogeneity of the environments. As shown in Fig. 4B, upon aTc
induction, two types of cell responses were observed: the fluorescence intensity
either changed significantly or almost remained the same. When we track cells in
real time, we can see that, some cells switch between bright and dim, while other
cells stay with similar brightness (Fig. 4B). The resulting fluorescence
distribution is thus bimodal and a fluorescence threshold can be defined for each
cell in its most probable induction state. The use of a microfluidic device,
coupled with cell tracking and fluorescence measurements, allows us to generate
fluorescence trajectories for a single cell on reasonable time scales
(\textasciitilde{}300 minutes) for a single trajectory. Based on this, we
collected 28 micro-colony movies and chose 163 fluorescence trajectories. We
observed that the trajectories of a single cell fluorescence fluctuated
significantly. We collected about 8200 fluorescence intensity data points
corresponding to the selected trajectories. Several representative trajectories
with significant fluctuations were shown to demonstrate the existence of two
states (From Figure 4A-B, Figure S8, Movie S1 and Movie S2).

\subsection*{Two Cell State Identifications by Hidden Markov Chain Modeling}

In order to explore the underlying mechanism of the bimodality, we collected the
statistics of the fluorescence intensity obtained from the trajectories. The
distribution of these intensities exhibits two peaks, suggesting that most of the
initial cells are either in a high expression state or in a low expression state
in their progeny. We then used a Hidden Markov Chain Model (HMM) (35) to fit the
real time trajectories and identify the cell states, and then simulate the
distribution of the fluorescence intensity (Figure 4C). To assign protein
expression states and the rates of inter-conversion between them, we performed
data fitting using the HMM. From the HMM analysis, we obtained a correlation
coefficient of 0.975 between the measured and simulated trajectories after
identifying the cell states and quantifying their switching rates. The simulated
distribution fits with the measured distribution well. From the HMM analysis, we
further determined the center positions of the peaks to be at 2.690 and at 2.933
in logarithm of fluorescence intensity. The variances of the individual peak
distributions are at 0.085 and at 0.080, respectively. For our system, the
probability in the high expression state is around 0.401, and we can also see
that the probability in the low expression state is around 0.599.

In the high expression state, the system will continue its behavior
with a probability of 0.963 (the switching or residence time will be discussed in
the next section). There is a small chance, with the probability of 0.037, to
switch to the low expression state from the high expression state. Meanwhile,
there is additionally a probability of 0.023 that the system will switch to the
high expression state from the low expression state, instead of remaining in the
low expression state.

\subsection*{The Average Residence Times of the Protein Expression States }

To estimate the average residence times of the protein expression state, we
distinguished the states from the trajectories using HMM analysis and calculated
the residence times of each state (Figure S11). For each trajectory, we counted
the total residence times and the number of the state changes. The average
residence times were calculated as the quotient of the total residence times and
the number of states changed.

The length of the test fluorescence trajectory is finite and limited. This may
lead to some errors in estimating the transition times. We take this into account
in determining the time scale of the transitions. The average residence time of
the high expression state is estimated to be about 92\textasciitilde{}103
minutes, and that of the low expression state is estimated to be about
151\textasciitilde{}182 minutes. The average residence time can be used to
quantify the switching time between two cell fates. Therefore, the switching time
from high (low) expression to low (high) expression can be estimated to be about
92\textasciitilde{}103 (151\textasciitilde{}182) minutes. Through fluctuations,
the bimodal distribution can be maintained in a dynamic balance between the high
expression ``on'' state and the low expression ``off'' state. When the inducer
concentration is fixed, the increasing number of proteins will promote the
inhibition probability of gene switching. Therefore, the cells in the high
expression state will have a tendency to migrate to the low expression state.
Conversely, the cells with low expressions will be more likely to move towards
the ``on'' state. Therefore, the cells in the low expression state will also have
a tendency to migrate to the high expression state.

\subsection*{Physical Origin of the Two Cell Fates }

Intuitively, from a molecular perspective, we know that the transcription
process is suppressed when the promoter site of the DNA is occupied by a
repressor (the gene is ``off''), and enhanced when the repressor is dissociated
from DNA (the gene is ``on''). When the inducer concentration is low, increasing
the inducer concentration will increase the binding of aTc to \textit{TetR} and
slow down the effective binding of \textit{TetR} to the promoter. This lessens
the chance of the genes being in an ``off state'' and conversely increases the
possibilities of the gene being at the ``on state'', resulting in higher
expressions. This explains the shift of the expression peak from low to high as
inducer concentration increases. When the inducer concentration further increases
to sufficiently high values, the chance of having free \textit{TetR} molecules
will be higher (comparable number of \textit{TetR} molecules to that of aTc
molecules) as a result of synthesis. More \textit{TetR} molecules will have
increased chances of binding to the promoter site and will therefore display more
repressive activity. This will lead to the emergence of the low expression peak
and therefore bimodal distribution of the copy number in mRNA and proteins.
Further increases of the inducer concentrations will lead to more free
\textit{TetR} molecules, with a resulting greater weighting of low, rather than
high, expression peaks. This explains the trend of expression peaks as seen in
Fig. 2A.

When the effective binding/unbinding is much faster compared to the
synthesis/degradation, the gene state changes rapidly. The interactions and the
mixings become stronger between the two gene states, and therefore also between
the two corresponding protein concentration peaks. For the self-repressor,
decreasing the effective binding (increasing the inducer concentrations in our
study) promotes the generation of more proteins which in turn shows greater
repressive activity. This leads to the high concentration peak moving towards a
lower concentration. On the other hand, increasing the binding (decreasing the
inducer concentrations in our study) represses the generation of the proteins,
and so fewer proteins produced bind effectively to DNA. This in turn promotes
production of \textit{TetR} molecules. It leads to the low concentration peak
moving outward towards a higher concentration. As a result, the two peaks from
the non-adiabatic limit (e.g. high aTc concentrations at 1400 ng/mL, slower
binding) meet in the adiabatic limit (e.g. lower aTc concentration at 1300 ng/mL,
faster binding) of the fast binding and merge into a single peak.

Gene switching is often rapid in bacterial cells. However, slow gene switching
controlled by regulatory proteins binding/unbinding to the promoters can also be
significant for gene expression dynamics. In eukaryotic cells and some
prokaryotic cells, binding/unbinding may be comparable to or even slower than the
corresponding synthesis and degradation due to epigenetic effects or complex
microenvironments. By studying how the introduction of the inducers effectively
weakens gene regulation in bacteria, we mimicked gene regulation dynamics in more
complex eukaryotic cells. Through increasing inducer concentrations, we achieved
effectively slower regulatory binding relative to synthesis and degradation. In
other words, the introduction of the inducers in the bacteria leads to an
additional time scale for regulatory binding. This mimicked the additional time
scales for regulatory binding from including the histone modifications and DNA
methylations in eukaryotic cells. This slower regulatory binding to inducers will
lead to prolonged times of genes being in the ``on state'' in addition to the
time spent in the ``off state'', originated from the fast binding without
inducers. As a result, both ``on'' and ``off'' states of genes may emerge. This
is the physical mechanism of bimodality. In other words, the fast regulatory
binding mimicked stronger interactions while the slow regulatory binding mimicked
the weaker interactions among genes. While stronger interactions give more
constraints to the system and therefore fewer degrees of freedom for the
expressions (single peak expression), the weaker interactions will constrain the
system less and therefore result in more degrees of freedom for the expressions
(e.g., double peak expressions).  Through the steady state and the real time
observations of the dynamics of the self-repressor in the experiments, we
observed the robust emergence of the bimodal gene expression distribution for the
self-repressor.

\subsection*{Stochastic Simulations of Bimodality }

We further explored the stochastic dynamics of self-regulative feedback genes
through a mathematical model, which can be used to explain and simulate the
experimental observations (Figure 3C). The mathematical model clarifies the
underlying mechanism of how bimodality emerges. Under faster regulation binding,
the self-repressor is forced to stay in the repressed state. This is because once
produced, the regulatory protein immediately binds to the gene and therefore
represses protein production. In our study, slower binding of the regulatory
protein to the gene is realized through the inducer binding to the regulatory
protein, which effectively blocks the ability of the protein to bind to the
promoter. Under slower regulatory binding, the self-repressor may function in two
different ways: it may bind to the DNA for some time and repress protein
production, or unbind from the DNA for some time, leading to increased protein
production. This generates two cell phenotypes. Furthermore, due to the intrinsic
statistical fluctuations of the number of proteins, there is a possibility of
switching between the high expression and low expression state. We have observed
such phenotypic switching in real time experiments. The simulation results are
consistent with the experimental observations.

On the other hand, the trajectories in Figure 4A and Fig. S8 showed comparable
growth rates in high expression state and in low expression state. It is possible
that high expression cells in our study have not reached the threshold for
significant metabolic burden to slow down the growth. The inhibition curves of
the different inducer concentrations in Fig. 3B and the dynamic balance by
intrinsic fluctuations also imply that the bimodality of the protein expression
distribution is not due to cell growth.

\section*{ Discussion}

For self-repressor gene network, even when the gene is fixed, there can still be
new cell phenotypes. Our study shows explicitly in this concrete gene circuit
that different cell fates can emerge not only from the changes in the genes (such
as mutations) but also from the changes in regulatory wirings or links through
microenvironments without altering the gene itself. In fact, even when the
topology of the wiring for the underlying gene regulatory network is fixed, there
is still a possibility of cell phenotypic changes due to the changes in the
regulation strengths induced by the environment. Furthermore, we observed both in
real time experiments and simulations that the cell phenotypes or fates can be
switched from one to the other. We also obtained the average time of this
switching which quantifies how difficult it is to communicate globally from one
cell fate to the other. Therefore, using real time trajectories, we determine
both the speed and the underlying processes of the cell fate
decision-making/phenotypic state-switching.

Epigenetic effects are often challenging to study in eukaryotic
cells. Our study in bacteria illustrates how the environments can influence the
cell fates and cell fate decision-making in a controllable way. The experiments
in bacteria are relatively easy and straightforward to perform and control. The
epigenetic and micro-environmental effects can be mimicked through the modulation
of inducers in our study. This is an advantage of our approach. We plan to apply
our method to a variety of core regulatory motifs and modules in the gene
networks to investigate how the environments or epigenetics influence the cell
fates and the cell fate decision-making processes.

\section*{Acknowledgments}

We thank H. Bujard and B.L.Wanner for providing pZE11 and CRIM vector systems
respectively, as well as for providing detailed information on their origins and
growth conditions. We would also like to thank X. S. Xie for providing \textit{E.
coli} SX4. This work was supported by the National Science Foundation (NSF) with
grant number 0947767, the National Science Foundation of China (NSFC) with grant
number 91430217 and the Ministry of Science and Technology (MOST) of China with
grant number 2016YFA0203200.

\section*{Author contributions}

Z.L. Jiang and L. Tian, X.N. Fang, Q. Z. Dong, and J. Wang contributed to the
experimental design. Z.L. Jiang, L. Tian, X. N. Fang and Q. Z. Dong conducted the
experiments. K. Zhang, Q. Liu, Z.L. Jiang and L. Tian, X. Fang, and J. Wang
contributed to data interpretation. Z.L. Jiang, L. Tian, X.N. Fang, E. K. Wang
and J. Wang contributed to writing and revising the manuscript.

%\section*{References}

\section*{Figure Captions}

\begin{figure}[!ht]
\centering
\includegraphics[width=1.0\textwidth]{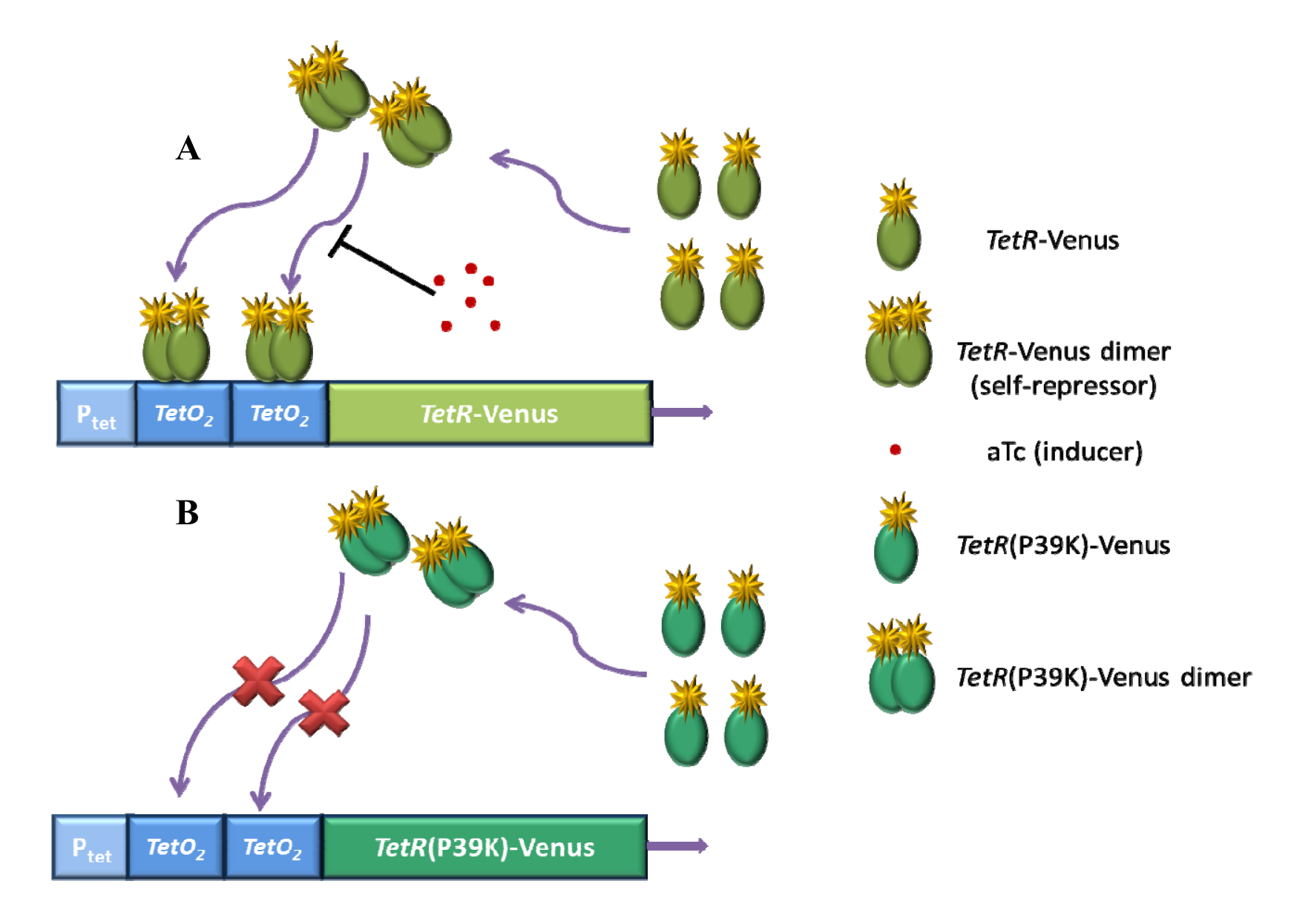}
\caption{Schematic illustrations of the self-repressing gene circuit
(MG::PR-8T) and the non-self-repressing gene circuit (MG::PR-8T-P39K).
(A) Two \textit{tet} operator sequences (T\textit{etO}$_{2}$) inserted
downstream of the P\textit{tet} promoter are bound by \textit{TetR}
self-repressor dimers. In the absence of aTcs (the inducers), \textit{TetR}-Venus
dimers bind to the operators. This interaction prevents the binding of RNA
polymerase, thereby inhibiting the \textit{TetR}-Venus fusion protein synthesis.
When aTcs diffuse into the cell, they bind to \textit{TetR}, inducing an
allosteric conformational change in the repressor protein which releases it from
DNA, allowing for the possibility of the gene being switched into the ``on''
state. All of these constitute a self-repressing gene circuit.
(B) The \textit{TetR}-P39K mutant is not capable of recognizing the operators
and is unable to repress the \textit{TetR}-Venus expression, constituting a
non-self-repression gene circuit.}\label{fig1}
\end{figure}

\begin{figure}[!ht]
\centering
\includegraphics[height=4.5in, width=5.0in]{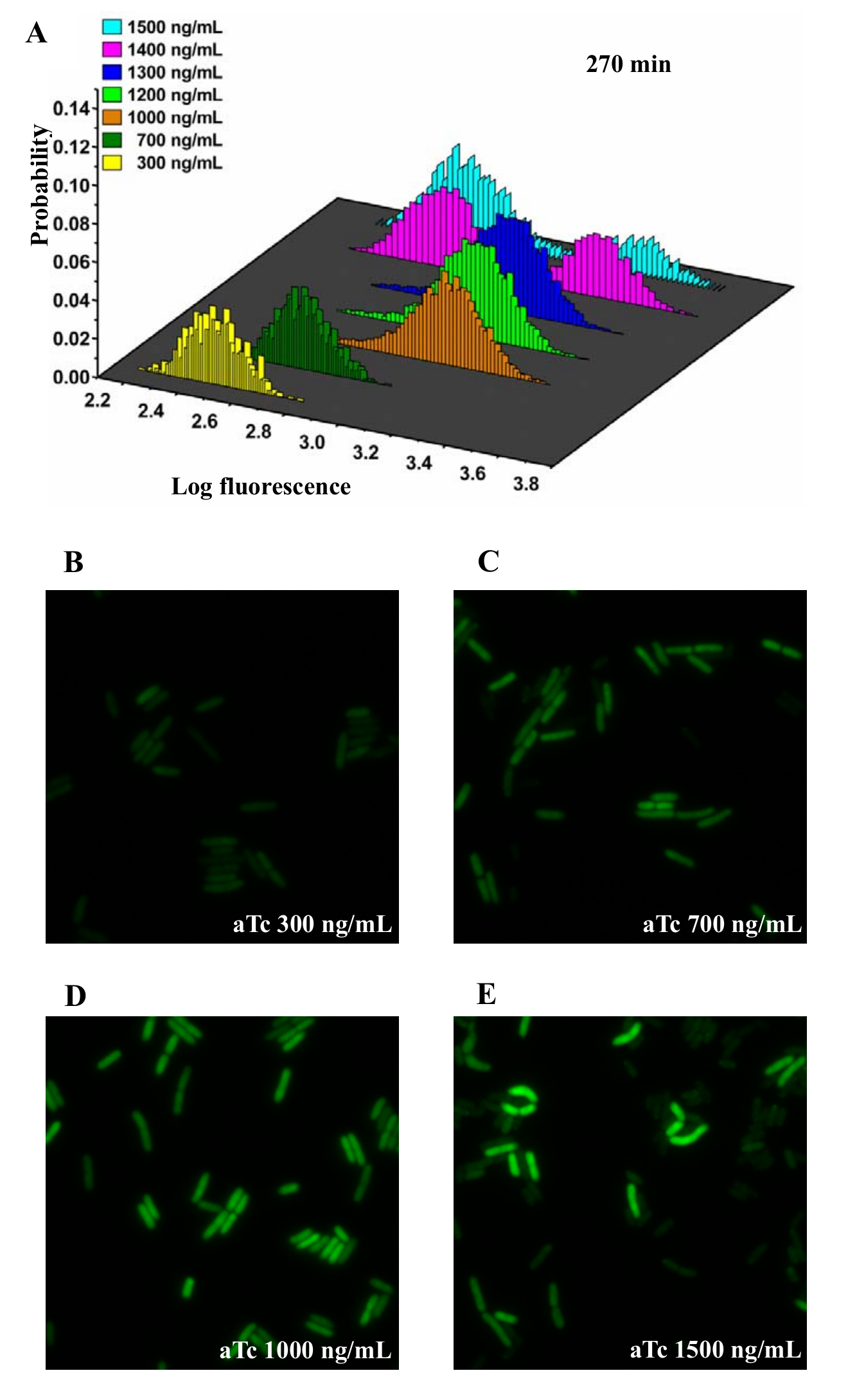}
\caption{Experimental expression distributions of the
self-repressing gene circuit (MG::PR-8T) at different aTc concentrations
observed under a microscope.
(A) In M9 media with the inducer concentrations ranging from 300 to 1500 ng/mL
of aTcs, the resulting steady state fluorescence distributions show that the
ratio of the populations of the bimodal fluorescence distributions depend on the
aTc concentration. Seven color histograms represent different inducer
concentrations.
(B, C, D, and E) Four representative fluorescence images at different
concentrations of aTcs (300, 700, 1000, and 1500 ng/mL) are selected.}\label{fig2}
\end{figure}

\begin{figure}[!ht]
\centering
\includegraphics[height=4.5in, width=5.0in]{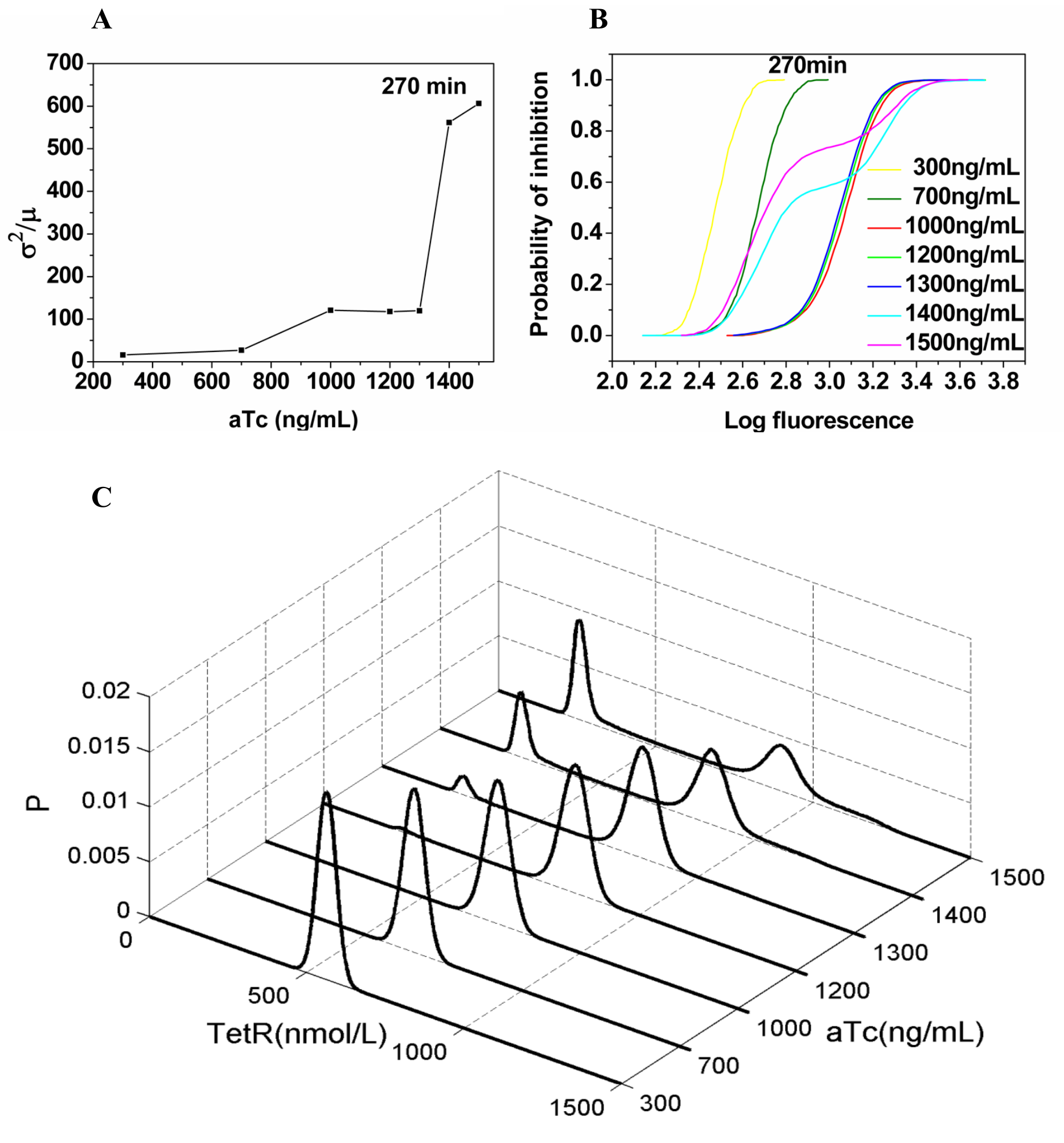}
\caption{The Fano factor curves and the probability of inhibition
curves of the self-repressing gene circuit.
(A) Dose-response of the Fano factor (\textit{F = $\sigma{}$$^{2}$/$\mu{}$}) of
the \textit{TetR}-Venus expression for the self-repressing gene circuit
(MG::PR-8T) at different inducer concentrations. The Fano factor is defined as
\textit{$\sigma{}$$^{2}$/$\mu{}$}, where \textit{$\sigma{}$$^{2}$} and
\textit{$\mu{}$} are the variance and the mean of the probability distribution.
(B) The probability of inhibition curves of the MG::PR-8T circuit at different
inducer concentrations. Seven color histograms represent different inducer
concentrations. The inhibition curves were obtained by the ratio of the cells
with a fluorescence intensity lower than a certain value to the number of the
total samples.
(C) The probability distribution of the \textit{TetR} proteins for the circuit
of MG::PR-8T with different concentrations of inducers from the stochastic
simulation model. P(n) (z axis) represents the probability distribution of the
TetR protein numbers (x axis), n at  different numbers of inducer (aTc) molecules
(y axis).}\label{fig3}
\end{figure}

\begin{figure}[!ht]
\centering
\includegraphics[height=4.0in, width=5.0in]{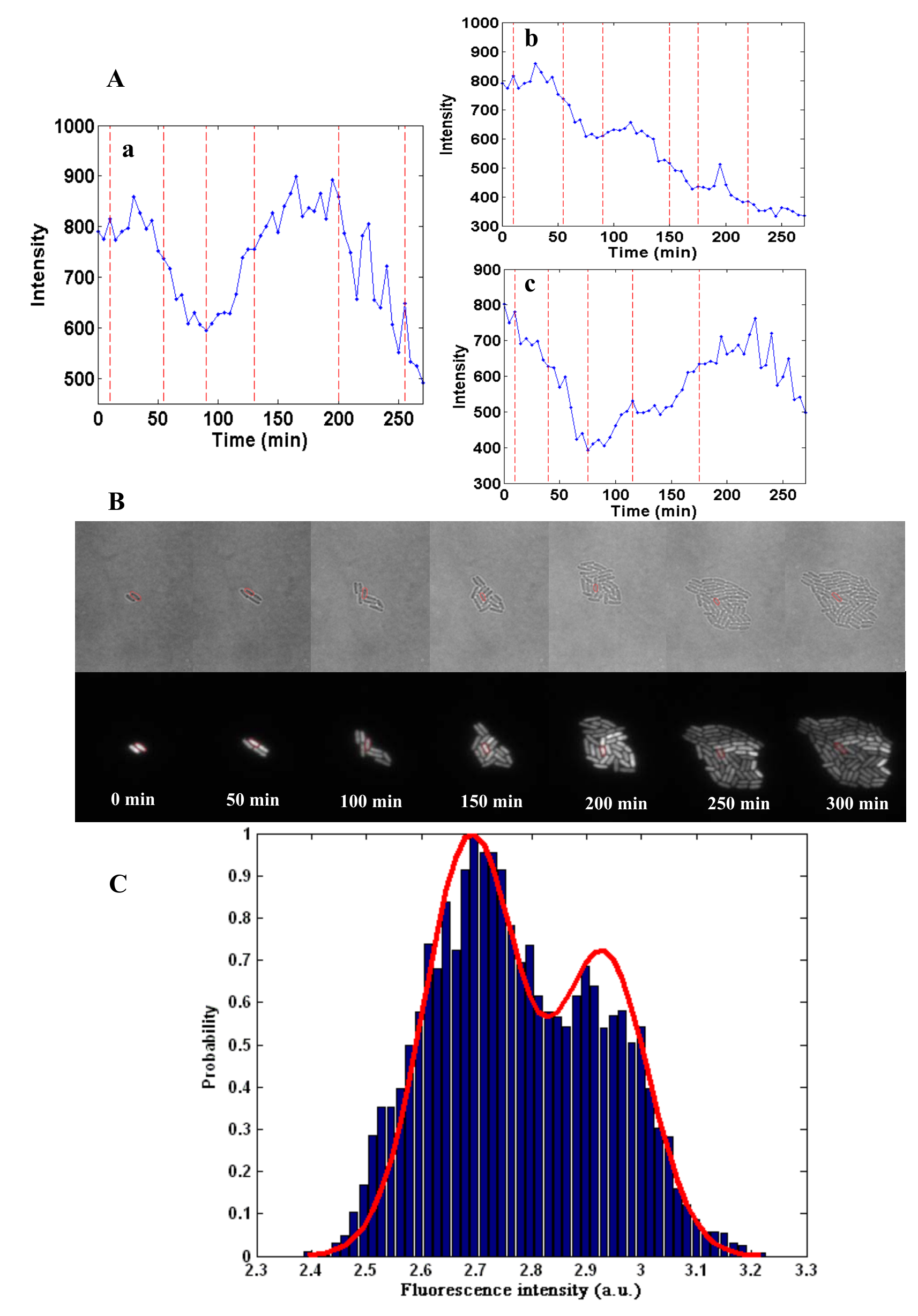}
\caption{ The mean fluorescence intensity distribution of the dynamical
trajectories for MG::PR-8T Single cell mean fluorescence intensities were captured every 5 minutes. 28
micro-colonies were tracked by time-lapse microscopy.
(A) Three representative single cell fluorescence trajectories induced by 1500
ng/mL aTcs. Points represent experimental fluorescence values. Red vertical
dashed lines demarcate cell divisions.
(B) The bright field and fluorescent field images of the corresponding
measurements in the time-lapse experiment. The cells corresponding to the
fluorescence trajectory in Fig. 4a are marked with red circles. The average of
bacteria mean fluorescence intensity is 556 and the average cell cycle time is 46
minutes in this micro-colony.
(C) The histogram gives the intensity distribution of the 163 single cell
fluorescence trajectories induced at 1500 ng/mL aTc collected from the time-lapse
experiments. The red solid curve is the fitted intensity distribution from HMM.}\label{fig4}
\end{figure}


\begin{thebibliography}{99}

\bibitem{1} Elowitz MB, Levine AJ, Siggia ED, \& Swain PS (2002) Stochastic
gene expression in a single cell. \textit{Science} 297(5584):1183-1186.

\bibitem{2} Kaern M, Elston TC, Blake WJ, \& Collins JJ (2005) Stochasticity
in gene expression: From theories to phenotypes. \textit{Nat Rev Genet}
6(6):451-464.

\bibitem{3} Becskei A, Kaufmann BB, \& van Oudenaarden A (2005) Contributions
of low molecule number and chromosomal positioning to stochastic gene expression.
\textit{Nat Genet} 37(9):937-944.

\bibitem{4} Swain PS, Elowitz MB, \& Siggia ED (2002) Intrinsic and extrinsic
contributions to stochasticity in gene expression. \textit{P Natl Acad Sci USA}
99(20):12795-12800.

\bibitem{5} Raj A \& van Oudenaarden A (2008) Nature, Nurture, or Chance:
Stochastic Gene Expression and Its Consequences. \textit{Cell} 135(2):216-226.

\bibitem{6} Qi H, Blanchard A, \& Lu T (2013) Engineered genetic information
processing circuits.~\textit{WIREs Syst Biol Med}~5(3): 273-287.

\bibitem{7} Graf T \& Enver T (2009) Forcing cells to change lineages.
\textit{Nature} 462(7273):587-594.

\bibitem{8} Fu WX, Ergun A, Lu T, Hill JA, Haxhinasto S, Fassett MS, Gazit R,
Adoro S, Glimcher L, Chan S, Kastner P, Rossi D, Collins JJ, Mathis D, \& Benoist
C (2012) A multiply redundant genetic switch 'locks in' the transcriptional
signature of regulatory T cells. \textit{Nat Immunol} 13(10):972-980.

\bibitem{9} Takahashi K \& Yamanaka S (2006) Induction of pluripotent stem
cells from mouse embryonic and adult fibroblast cultures by defined factors.
\textit{Cell} 126(4):663-676.

\bibitem{10} Yamanaka S \& Blau HM (2010) Nuclear reprogramming to a
pluripotent state by three approaches. \textit{Nature} 465(7299):704-712.

\bibitem{11} Yamanaka S (2009) Elite and stochastic models for induced
pluripotent stem cell generation. \textit{Nature} 460(7251):49-52.

\bibitem{12} Wang J, Xu L, \& Wang EK (2008) Potential landscape and flux
framework of nonequilibrium networks: Robustness, dissipation, and coherence of
biochemical oscillations. \textit{P Natl Acad Sci USA} 105(34):12271-12276.

\bibitem{13} Wang J, Zhang K, Xu L, \& Wang E (2011) Quantifying the
Waddington landscape and biological paths for development and differentiation.
\textit{P Natl Acad Sci USA} 108(20):8257-8262.

\bibitem{14} Wang J (2015) Landscape and flux theory of non-equilibrium
dynamical systems with application to biology. \textit{Adv Phys} 64(1):1-137.

\bibitem{15} Sasai M \& Wolynes PG (2003) Stochastic gene expression as a
many-body problem. \textit{P Natl Acad Sci USA} 100(5):2374-2379.

\bibitem{16} Balazsi G, van Oudenaarden A, \& Collins JJ (2011) Cellular
Decision Making and Biological Noise: From Microbes to Mammals. \textit{Cell}
144(6):910-925.

\bibitem{17} Liao C, \& Lu T (2013) A minimal transcriptional controlling
network of regulatory T cell development.~\textit{J Phys Chem B}
117(42):12995-13004.

\bibitem{18} Nevozhay D, Adams R, Murphy K, Josic K, Bal\'{a}zsi G (2009)~Negative
autoregulation linearizes the dose response and suppresses the heterogeneity of
gene expression.~\textit{P Natl Acad Sci USA} 106(13):5123-5128

\bibitem{19} Koga M, Matsuda M, Kawamura T, Sogo T, Shigeno A, Nishida E, \&
Ebisuya M (2014) Foxd1 is a mediator and indicator of the cell reprogramming
process. \textit{Nat Commun} 5:3197.

\bibitem{20} Buganim Y, Faddah DA, \& Jaenisch R (2013) Mechanisms and models
of somatic cell reprogramming. \textit{Nat Rev Genet} 14(6):427-439.

\bibitem{21} Pijnappel WWMP, Esch D, Baltissen MPA, Wu GM, Mischerikow N,
Bergsma AJ, van der Wal E, Han DW, vom Bruch H, Moritz S, Lijnzaad P, Altelaar
AFM, Sameith K, Zaehres H, Heck AJR, Holstege FCP, Sch\"{o}ler HR, \& Timmers HTM
(2013) A central role for TFIID in the pluripotent transcription circuitry.
\textit{Nature} 495(7442):516-519.

\bibitem{22} Singh A \& Weinberger LS (2009) Stochastic gene expression as a
molecular switch for viral latency. \textit{Curr Opin Microbiol} 12(4):460-466.

\bibitem{23} Yu J, Xiao J, Ren XJ, Lao KQ, \& Xie XS (2006) Probing gene
expression in live cells, one protein molecule at a time. \textit{Science}
311(5767):1600-1603.

\bibitem{24} Thattai M \& van Oudenaarden A (2001) Intrinsic noise in gene
regulatory networks. \textit{P Natl Acad Sci USA} 98(15):8614-8619.

\bibitem{25} Becskei A \& Serrano L (2000) Engineering stability in gene
networks by autoregulation. \textit{Nature} 405(6786):590-593.

\bibitem{26} Lutz R \& Bujard H (1997) Independent and tight regulation of
transcriptional units in Escherichia coli via the LacR/O, the TetR/O and
AraC/I-1-I-2 regulatory elements. \textit{Nucleic Acids Res} 25(6):1203-1210.

\bibitem{27} Hasty J, McMillen D, \& Collins JJ (2002) Engineered gene
circuits. \textit{Nature} 420(6912):224-230.

\bibitem{28} Austin DW, Allen MS, McCollum JM, Dar RD, Wilgus JR, Sayler GS,
Samatova NF, Cox CD, \& Simpson ML (2006) Gene network shaping of inherent noise
spectra. \textit{Nature} 439(7076):608-611.

\bibitem{29} Maithreye R, Sarkar RR, Parnaik VK, \& Sinha S (2008)
Delay-Induced Transient Increase and Heterogeneity in Gene Expression in
Negatively Auto-Regulated Gene Circuits. \textit{Plos One} 3(8):e2972.

\bibitem{30} Guet CC, Elowitz MB, Hsing WH, \& Leibler S (2002) Combinatorial
synthesis of genetic networks. \textit{Science} 296(5572):1466-1470.

\bibitem{31} Kepler TB \& Elston TC (2001) Stochasticity in transcriptional
regulation: Origins, consequences, and mathematical representations.
\textit{Biophys J} 81(6):3116-3136.

\bibitem{32} Hornos JEM, Schultz D, Innocentini GCP, Wang J, Walczak AM,
Onuchic JN, \& Wolynes PG (2005) Self-regulating gene: an exact solution.
\textit{Physical review. E, Statistical, nonlinear, and soft matter physics} 72(5
Pt 1):051907.

\bibitem{33} Feng HD, Han B, \& Wang J (2011) Adiabatic and Non-Adiabatic
Non-Equilibrium Stochastic Dynamics of Single Regulating Genes. \textit{J Phys
Chem B} 115(5):1254-1261.

\bibitem{34} Fano U (1947) Ionization yield of radiations. II. The
fluctuations of the number of ions. \textit{Phys. Rev} 72(1):26-29.

\bibitem{35} Baum LE \& Petrie T (1966) Statistical Inference for
Probabilistic Functions Of Finite State Markov Chains. \textit{Ann Math Stat}
37(6):1554-1563.

\end{thebibliography}
\end{document}